\documentstyle[twocolumn,aps,psfig]{revtex}
\begin{document}
%\eqnobysec   %JPB
%\jl{2}    %JPB
%\twocolumn[
\draft    %JPB

\title{Curve crossing in linear potential grids: the quasidegeneracy
approximation}

\author{V. A. Yurovsky and A. Ben-Reuven}

\address{School of Chemistry, Tel Aviv University, 69978
Tel Aviv, Israel}

\date{\today}
\maketitle    %JPB
%\widetext
\begin{abstract}

The quasidegeneracy approximation [V. A. Yurovsky, A.
Ben-Reuven, P. S. Julienne, and Y. B. Band, J. Phys. B {\bf 32},
1845 (1999)] is used here to evaluate transition amplitudes
for the problem of curve crossing in linear potential grids
involving two sets of parallel potentials. The approximation
describes phenomena, such as counterintuitive transitions and
saturation (incomplete population transfer), not predictable
by the assumption of independent crossings. Also, a new kind
of oscillations due to quantum interference (different from
the well-known St\"uckelberg oscillations) is disclosed, and its
nature discussed. The approximation can find applications in
many fields of physics, where multistate curve crossing
problems occur. \end{abstract}

%]
\narrowtext

\section{Introduction}

The concept of curve crossing has many applications in the study
of atomic collisions \cite{NU84,N99,ChildMCT,ChildSM,NWJ97,YB97,FC00},
excitations of atoms and molecules by nonstationary fields
\cite{Shore90,VS99,A84,HP94}, Bose-Einstein condensates
\cite{YBJW99,MJT99}, and solid state physics \cite{U97,U98}. Typical
curve crossing problems are generally divided into two classes,
$R$-dependent and $t$-dependent ones. The description of inelastic
collisions, for example, involves crossings of coordinate-dependent
potentials, and can therefore be treated as an $R$-dependent problem,
described by a set of coupled second-order stationary Schr\"odinger
equations. By the use of a common-trajectory approximation
\cite{NU84}, $R$-dependent problems can be reduced to $t$-dependent
 ones.
The latter class also naturally appears in the description of
transitions due to non-stationary fields. Typical $t$-dependent
 problems
involve the crossing of time-varying potentials, and their description
requires a set of coupled first-order non-stationary Schr\"odinger
equations.

Curve crossing problems are usually solved by using semiclassical
approximations. Two-state crossing is described by the Landau-Zener
(LZ) formula \cite{L32,Z32} (or one of its various modifications
\cite{NZ96}), while multistate crossing is treated as a sequence of
independent two-state crossings. Although semiclassical approaches are
satisfactory for many applications (see, e.g. \cite{YB97}), they fail
to describe certain effects found recently in experiments, numerical
calculations, and analytically-soluble potential models
\cite{NWJ97,YBJB99,N62,B72,YB99}.

The LZ formula forms an exact solution of the problem involving
the crossing of two linear potentials infinitely diverging on both
asymptotes. It gives good results even when local perturbations in
the vicinity of the crossing are taken into account \cite{VS99}.
However, if the potentials retain a finite potential gap on an
asymptote \cite{N62}, have singularities \cite{B72,YB99}, or are
truncated \cite{VG96}, the transition probabilities deviate
essentially from the LZ formula. Some interesting effects appear also
in state crossing involving Bose-Einstein condensates, described by
the nonlinear Gross-Pitaevskii equations (see \cite{YBJW99}), instead
of the linear Schr\"odinger equations.

\begin{figure}
\psfig{clip=,figure=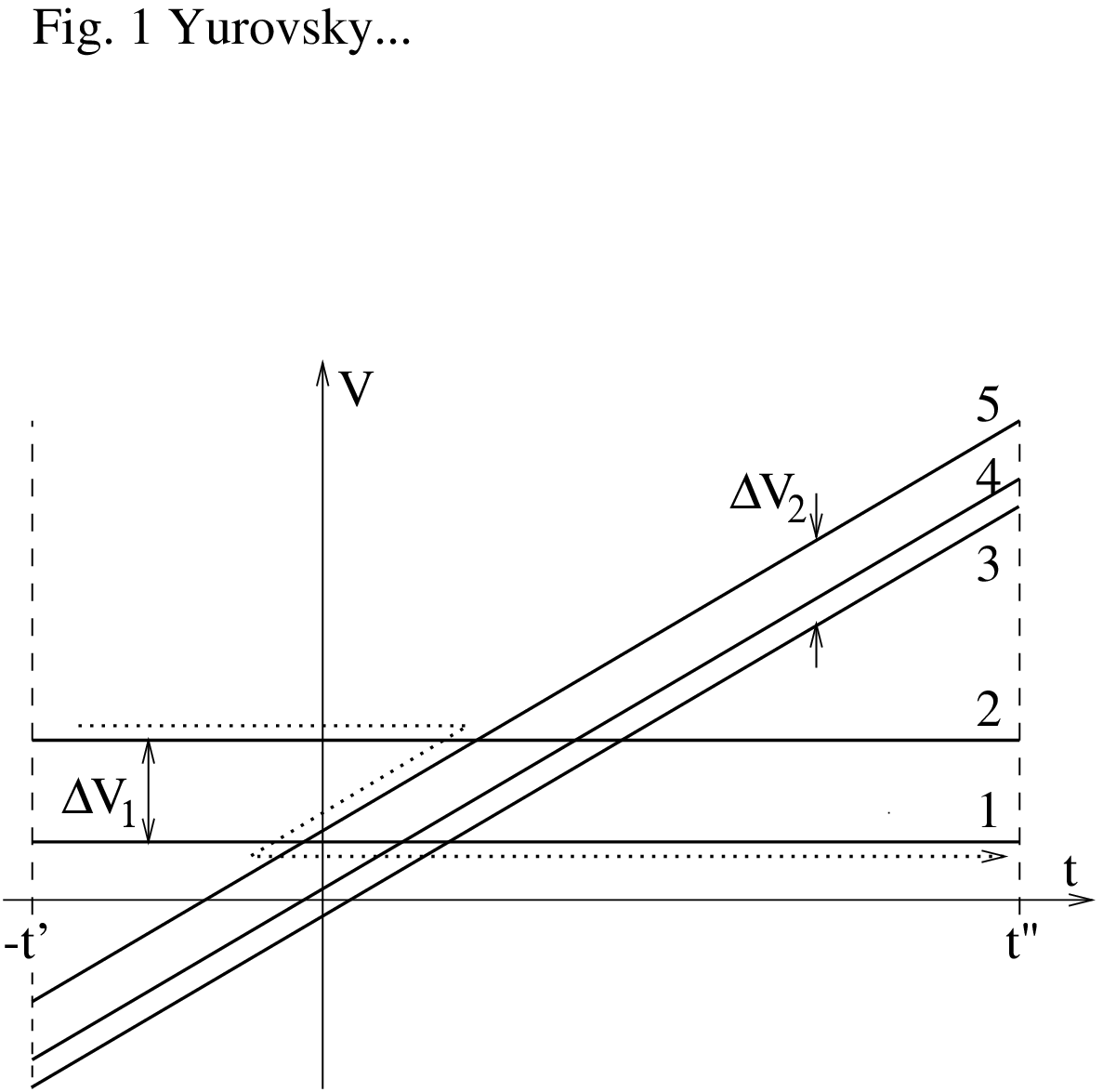,width=3.375in}

\caption{Schematic illustration of a truncated linear grid,
involving $n_{1}=2$ horizontal potentials and $n_{2}=3$ slanted
 potentials. The
broken dotted arrow shows a counterintuitive transition. The numbers
denote the states to which the potentials correspond.} \label{FigPot}

\end{figure}

The treatment of multistate curve crossing as a sequence of
independent two-state crossings, commonly used in semiclassical
approaches, fails to describe ``counterintuitive'' transitions (see
\cite{NWJ97,YBJB99}), in which the second crossing precedes the first
one. In the case of a $t$-dependent linear grid (see
\cite{HP94,U97,U98,BE93,Ostrovsky}) consisting of two sets of
mutually parallel potentials (see Fig.\ \ref{FigPot}),
counterintuitive transitions are exactly forbidden if the problem is
defined on the infinite time interval ($t^\prime \rightarrow \infty ,
 t^{\prime\prime}\rightarrow -\infty $), as has been
shown in Ref.\ \cite{BE93}. However, the potentials in the grid
infinitely diverge on both asymptotes, which is unphysical. We
consider here a truncated linear grid, defined on a finite time
interval $\left\lbrack -t^\prime , t^{\prime\prime}\right\rbrack $.
 Such a truncation is relevant for application to
transitions in time-dependent fields (e.g., single electronics
\cite{U97,U98}), since the field variation is actually finite.

The problem is studied here with the help of the
``quasidegeneracy'' approximation, introduced in \cite{YBJW99} for the
case when one of the sets consists of only one potential. This
approximation treats a non-degenerate system with small potential gaps
as a perturbed degenerate system (see \cite{YB98}). A special form of
the quasidegeneracy approximation was used also in Ref.\ \cite{U97}.
This form is applicable to a linear grid consisting of two potentials
in each set, with equal couplings between all states belonging to
different sets, while only one of the sets is quasidegenerate. The
method of Ref.\ \cite{U97} is actually a simplified form of the method
of Ref.\ \cite{YBJB99}. A linear grid in which one set of parallel
potentials is exactly degenerate was considered also in Ref.\
\cite{A84}, by using a method different from the quasidegeneracy
approximation.

The quasidegeneracy approximation is generalized here to the case
of a truncated linear grid with arbitrary number of potentials in both
sets. In Sec.\ \ref{SecTran} we introduce a ``decoupling''
transformation, which approximately transforms the problem to a set of
parallel two-state crossings. The transition amplitudes are calculated
in Sec.\ \ref{SecAmp}, and applicability criteria are presented in
Sec.\ \ref{SecCrit}. Results are shown and discussed in Sec.\
\ref{SecDisc}. A few preliminary results of this work have been
presented in \cite{YB00}.

\section{Decoupling transformation} \label{SecTran}

Let us consider two sets of mutually parallel linear potentials.
This problem can be easily reduced by a gauge transformation to the
case of a set of horizontal potentials $V_{j}$  ($j=1,\ldots  ,n_{1}$
) crossed by a
set of slanted parallel linear potentials $V_{n_1+k}+\beta t$
($k=1,\ldots  ,n_{2}$) (see
Fig.\ \ref{FigPot}). The interactions between the states within each
set of parallel potentials can be eliminated by a unitary
transformation. Therefore, without loss of generality, we can describe
the problem by the following system of coupled equations for the
expansion coefficients $\varphi _{j}\left( t\right) $,
\begin{eqnarray}
&&i{\partial \varphi { } _{j}\over \partial t}=V_{j}\varphi _{j}
+\sum\limits^{n{ } _{2}}_{k=1}g_{jk}\varphi _{n_1+k} ,\qquad 1\le
 j\le n_{1} \nonumber
\\
&&{} \label{phieq}
\\
&&i{\partial \varphi { } _{n_1+k}\over \partial t}=\left( V_{n_1+k}
+\beta t\right) \varphi _{n_1+k}+\sum\limits^{n{ } _{1}}_{j=1}g^{
*}_{jk}\varphi _{j} , \quad 1\le k\le n_{2} \nonumber
\end{eqnarray}
(using a system of units in which $\hbar =1$). The only non-vanishing
coupling coefficients $g_{jk}$  involve pairs of crossed potentials.
 The
problem is defined here on the finite time-interval $-t^\prime \le
 t\le t^{\prime\prime}$.

The special case of $n_{2}=1$ has been considered in \cite{YBJB99}, by
using a quasidegeneracy approximation. In order to generalize this
approximation, let us perform a singular value decomposition (SVD) for
the coupling matrix $g_{jk}$, of the form
\begin{equation}
g_{jk}=\sum\limits^{n}_{l=1}X^{*}_{lj}g_{l}Y_{lk},\qquad n\le
 \min\left( n_{1},n_{2}\right)  .
\end{equation}
This decomposition is well known in the theory of spline
approximations (see, e. g., \cite{SVD}). The two matrices with
elements $X_{lj}$  and $Y_{lk}$  are unitary, and their rows are the
eigenvectors of the quadratic matrices formed by products of the
 $g_{jk}$
and their hermitian conjugates:
\begin{eqnarray}
&&\sum\limits^{}_{k,j}g^{*}_{j^\prime k}g_{jk}X_{lj}=|g_{l}
|^{2}X_{lj^\prime } , \nonumber
\\
&&\sum\limits^{}_{k,j}g_{jk^\prime }g^{*}_{jk}Y_{lk}=|g_{l}
|^{2}Y_{lk^\prime } . \nonumber
\end{eqnarray}
A transformation of the expansion coefficients $\varphi _{j}\left(
 t\right) $ using the
matrices $X_{lj}$  and $Y_{lk}$,
\begin{equation}
a_{l}\left( t\right) =\sum\limits^{n{ } _{1}}_{j=1}X_{lj}\varphi
 _{j}\left( t\right)  ,\qquad b_{l}\left( t\right) =\sum\limits^{n{ }
 _{2}}_{k=1}Y_{lk}\varphi _{n_1+k}\left( t\right)  , \label{ab}
\end{equation}
leads to a new system of coupled equations
\begin{eqnarray}
&&i{\partial a{ } _{l}\over \partial t}=V^{\left( a\right)
 }_{ll}a_{l}+g_{l}b_{l}+\sum\limits^{}_{l^\prime \neq l}V^{\left(
 a\right) }_{ll^\prime }a_{l^\prime } ,\qquad 1\le l\le n \nonumber
\\
&&i{\partial b{ } _{l}\over \partial t}=\left( V^{\left( b\right)
 }_{ll}+\beta t\right) b_{l}+g_{l}a_{l}+\sum\limits^{}_{l^\prime \neq
 l}V^{\left( b\right) }_{ll^\prime }b_{l^\prime } ,\quad 1\le l\le n
 \nonumber
\\
&& {} \label{trde}
\\
&&i{\partial a{ } _{l}\over \partial t}=V^{\left( a\right)
 }_{ll}a_{l}+\sum\limits^{}_{l^\prime \neq l}V^{\left( a\right)
 }_{ll^\prime }a_{l^\prime } ,\qquad n+1\le l\le n_{1} \nonumber
\\
&&i{\partial b{ } _{l}\over \partial t}=\left( V^{\left( b\right)
 }_{ll}+\beta t\right) b_{l}+\sum\limits^{}_{l^\prime \neq l}V^{\left
( b\right) }_{ll^\prime }b_{l^\prime } ,\quad n+1\le l\le n_{2} ,
 \nonumber
\end{eqnarray}
in which
\begin{equation}
V^{\left( a\right) }_{ll^\prime }=\sum\limits^{n{ }
 _{1}}_{j=1}X_{lj}V_{j}X^{*}_{l^\prime j} ,\qquad V^{\left( b\right)
 }_{ll^\prime }=\sum\limits^{n{ } _{2}}_{k=1}Y_{lk}V_{n_1+k}Y^{
*}_{l^\prime k} .
\end{equation}
Given a matrix $g_{jk}$,  its SVD is not unique, and may be chosen in
such a way that its singular values $g_{l}$  are real and non-negative,
and the non-diagonal potential elements $V^{\left( a\right) }_{ll^\prime
 }$  and $V^{\left( b\right) }_{ll^\prime }$  vanish when $l>n$
and $l^\prime >n$.

When both parallel sets of potentials are degenerate, the
matrices $V^{\left( a\right) }_{ll^\prime }$  and $V^{\left( b\right)
 }_{ll^\prime }$  are diagonal, and the system (\ref{trde})
describes a set of $n$ independent pairs of crossing potentials,
 $n_{1}-n$
separate horizontal potentials (not coupled to other channels), and
$n_{2}-n$ separate slanted potentials. Since the transformation
(\ref{ab})
partially eliminates the coupling between the states, hereafter it is
called the ''decoupling transformation''. The channels described by
coefficients $a$ and $b$ will be called the ``decoupled channels''.

In the non-degenerate case the non-diagonal elements of $V^{\left(
 a\right) }_{ll^\prime }$  and
$V^{\left( b\right) }_{ll^\prime }$  lead to transitions between the
 decoupled channels. However, the
magnitudes of these non-diagonal elements are bounded by the
inequalities
\begin{eqnarray}
&&\sum\limits^{}_{l\neq l^\prime }|V^{\left( a\right) }_{ll^\prime }
|^{2}=\sum\limits^{n{ } _{1}}_{j=1}V^{2}_{j}-\sum\limits^{n{ }
 _{1}}_{l=1}\left( V^{\left( a\right) }_{ll}\right) ^{2}\le {1\over
 4}n_{1}\Delta V^{2}_{1} , \nonumber
\\
&& \label{Vnd}
\\
&&\sum\limits^{}_{l\neq l^\prime }|V^{\left( b\right) }_{ll^\prime }
|^{2}=\sum\limits^{n{ } _{2}}_{k=1}V^{2}_{n_1+k}-\sum\limits^{n{ }
 _{2}}_{l=1}\left( V^{\left( b\right) }_{ll}\right) ^{2}\le {1\over
 4}n_{2}\Delta V^{2}_{2} , \nonumber
\end{eqnarray}
where the bandwidths of the potential sets are defined as
\begin{equation}
\Delta V_{1}=V_{n_1}-V_{1} ,\qquad \Delta V_{2}=V_{n_1+n_2}-V_{n_1+1}
 . \label{DV12}
\end{equation}
Therefore, these transitions are negligible if the bandwidths
of the two potential sets are small enough. (Appropriate
applicability criteria are presented in Sec.\ \ref{SecCrit} below.)
Neglecting the non-diagonal elements of $V^{\left( a\right)
 }_{ll^\prime }$  and $V^{\left( b\right) }_{ll^\prime }$, we obtain a
zero-order-approximation system of equations for $a_{l}\left( t\right
) $ and $b_{l}\left( t\right) $,
\begin{mathletters} \label{zode}
\begin{eqnarray}
&&i{\partial a{ } ^{\left( 0\right) }_{l}\over \partial t}=V^{\left(
 a\right) }_{ll}a^{\left( 0\right) }_{l}+g_{l}b^{\left( 0\right)
 }_{l} ,\qquad 1\le l\le n \label{zodeac}
\\
&&i{\partial b{ } ^{\left( 0\right) }_{l}\over \partial t}=\left(
 V^{\left( b\right) }_{ll}+\beta t\right) b^{\left( 0\right) }_{l}
+g_{l}a^{\left( 0\right) }_{l} ,\quad 1\le l\le n \label{zodebc}
\\
&&i{\partial a{ } ^{\left( 0\right) }_{l}\over \partial t}=V^{\left(
 a\right) }_{ll}a^{\left( 0\right) }_{l} ,\qquad n+1\le l\le n_{1}
 \label{zodeau}
\\
&&i{\partial b{ } ^{\left( 0\right) }_{l}\over \partial t}=\left(
 V^{\left( b\right) }_{ll}+\beta t\right) b^{\left( 0\right) }_{l}
 ,\quad n+1\le l\le n_{2} , \label{zodebu}
\end{eqnarray}
\end{mathletters} which describes the same set of decoupled
channels as the one that prevails in the case of degenerate
potentials.

Given an arbitrary matrix $g_{jk}$,  the transformation matrices
 $X_{lj}$
and $Y_{lk}$  cannot be generally expressed in an analytical form.
Nevertheless, analytical expressions can be obtained in the specific
case of a separable matrix $g_{jk}=\xi ^{*}_{j}\eta _{k}$. In this
 case, one of the rows
(the first, for definiteness) has the form
\begin{equation}
X_{1j}=\left( \sum\limits^{n{ } _{1}}_{j^\prime =1}|\xi _{j^\prime }
|^{2}\right) ^{-1/2}\xi _{j} ,\quad Y_{1k}=\left( \sum\limits^{n{ }
 _{2}}_{k^\prime =1}|\eta _{k^\prime }|^{2}\right) ^{-1/2}\eta _{k} ,
\end{equation}
and the other rows are orthogonal to the first one. The singular
values can then be written as
\begin{equation}
g_{l}=\left( \sum\limits^{n{ } _{1}}_{j^\prime =1}|\xi _{j^\prime }
|^{2}\sum\limits^{n{ } _{2}}_{k^\prime =1}|\eta _{k^\prime }
|^{2}\right) ^{1/2}\delta _{l1} .
\end{equation}
In this case, $n=1$ and the transformed system consists of one pair
of coupled potentials, together with $n_{1}-1$ horizontal, and $n_{2}
-1$
slanted, separate potentials.

In the case of equal couplings $g_{jk}=g$ (independent of $j$ and $k$
),
$X_{1j}=n^{-1/2}_{1}$, $Y_{1j}=n^{-1/2}_{2}$, and $g_{l}=\left(
 n_{1}n_{2}\right) ^{1/2}g\delta _{l1}$. The opposite situation (in
which $g_{l}$  is independent of $l$) takes place in the case in
 which the
coupling matrix $g_{jk}$  is proportional to a unitary matrix.

\section{Transition amplitudes}  \label{SecAmp}

The zero-order equations (\ref{zodeau}) and (\ref{zodebu})
representing separate channels have the simple analytical solutions
\begin{eqnarray}
a^{\left( 0\right) }_{l}\left( t^{\prime\prime}\right) =&&a^{\left(
 0\right) }_{l}\left( -t^\prime \right) \exp\left( -i V^{\left(
 a\right) }_{ll}\left( t^\prime +t^{\prime\prime}\right) \right)  ,
 \nonumber
\\
&& \label{abu}
\\
b^{\left( 0\right) }_{l}\left( t^{\prime\prime}\right) =&&b^{\left(
 0\right) }_{l}\left( -t^\prime \right)  \nonumber
\\
&&\times \exp\left( -i V^{\left( b\right) }_{ll}\left( t^\prime
+t^{\prime\prime}\right) -i\beta \left( t^{\prime\prime}{}^{2}
-t^\prime {}^{2}\right) /2\right)  . \nonumber
\end{eqnarray}
The remaining equations (\ref{zodeac}) and (\ref{zodebc})
represent a set of $n$ two-state linear curve-crossing problems. In
 the
limit $t^\prime \rightarrow \infty $, $t^{\prime\prime}\rightarrow
 \infty $ the transition amplitude in each of these systems is
given by the LZ formula. However, the solution of this problem
converges to the asymptotic limit very slowly. We shall therefore use
the exact solution of the linear two-state curve crossing problem,
known since the pioneering work of Zener \cite{Z32}. The two
independent solutions $A_{ml}\left( t\right) $, $B_{ml}\left( t\right
) $, with $m=1,2$, can be expressed in
terms of the confluent hypergeometric function $_{1}F_{1}$  (see
\cite{Abramovitz}) as
\begin{eqnarray}
A_{1l}\left( t\right) =&&{}_{1}F_{1}\left( -{i\over 2}\lambda _{l},
 {1\over 2},-{i\over 2}\beta \left( t-t_{l}\right) ^{2}\right)
 \exp\left( -i V^{\left( a\right) }_{ll}t\right)  , \nonumber
\\
A_{2l}\left( t\right) =&&\left( t-t_{l}\right) {}_{1}F_{1}\left(
 {1\over 2}-{i\over 2}\lambda _{l}, {3\over 2},-{i\over 2}\beta \left
( t-t_{l}\right) ^{2}\right)  \nonumber
\\
&&\times \exp\left( -i V^{\left( a\right) }_{ll}t\right)  ,
 \label{fundsol}
\\
B_{ml}\left( t\right) &&={i\over g{ } _{l}}{\partial A_{ml}\left(
 t\right) \over \partial t}-{V{ } ^{\left( a\right) }_{ll}\over g{ }
 _{l}}A_{ml}\left( t\right)  , \nonumber
\end{eqnarray}
where
\begin{equation}
\lambda _{l}=g^{2}_{l}/\beta  ,\qquad t_{l}=\left( V^{\left( a\right)
 }_{ll}-V^{\left( b\right) }_{ll}\right) /\beta  \label{lambda}
\end{equation}
are, respectively, the LZ exponent for the two-state crossing and
the position of the crossing point on the time scale.

The transition matrix $S^{\left( l\right) }$, connecting the
 coefficients $a^{\left( 0\right) }_{l}$,
$b^{\left( 0\right) }_{l}$  at the boundaries $t^{\prime\prime}$  and
 $-t^\prime $  as
\begin{eqnarray}
&&a^{\left( 0\right) }_{l}\left( t^{\prime\prime}\right) =S^{\left(
 l\right) }_{aa}a^{\left( 0\right) }_{l}\left( -t^\prime \right)
+S^{\left( l\right) }_{ab}b^{\left( 0\right) }_{l}\left( -t^\prime
 \right)  , \nonumber
\\
&&{}
\\
&&b^{\left( 0\right) }_{l}\left( t^{\prime\prime}\right) =S^{\left(
 l\right) }_{ba}a^{\left( 0\right) }_{l}\left( -t^\prime \right)
+S^{\left( l\right) }_{bb}b^{\left( 0\right) }_{l}\left( -t^\prime
 \right)  , \nonumber
\end{eqnarray}
can be expressed in terms of the fundamental solutions
(\ref{fundsol}) in the form
\begin{eqnarray}
&&S^{\left( l\right) }_{aa}=\left( A_{1l}\left(
 t^{\prime\prime}\right) B_{2l}\left( -t^\prime \right) -A_{2l}\left(
 t^{\prime\prime}\right) B_{1l}\left( -t^\prime \right) \right)
/D_{l} , \nonumber
\\
&&S^{\left( l\right) }_{ab}=\left( -A_{1l}\left(
 t^{\prime\prime}\right) A_{2l}\left( -t^\prime \right) +A_{2l}\left(
 t^{\prime\prime}\right) A_{1l}\left( -t^\prime \right) \right)
/D_{l} , \nonumber
\\
&&{} \label{Sl}
\\
&&S^{\left( l\right) }_{ba}=\left( B_{1l}\left(
 t^{\prime\prime}\right) B_{2l}\left( -t^\prime \right) -B_{2l}\left(
 t^{\prime\prime}\right) B_{1l}\left( -t^\prime \right) \right)
/D_{l} , \nonumber
\\
&&S^{\left( l\right) }_{bb}=\left( -B_{1l}\left(
 t^{\prime\prime}\right) A_{2l}\left( -t^\prime \right) -B_{2l}\left(
 t^{\prime\prime}\right) A_{1l}\left( -t^\prime \right) \right)
/D_{l} , \nonumber
\end{eqnarray}
where
\begin{equation}
D_{l}=A_{1l}\left( -t^\prime \right) B_{2l}\left( -t^\prime \right)
-A_{2l}\left( -t^\prime \right) B_{1l}\left( -t^\prime \right)  .
\end{equation}
In our numerical calculations we expressed the confluent
hypergeometric functions in terms of columbic wavefunctions, using the
algorithm of \cite{TB85} for their evaluation.

When the original representation  (\ref{phieq}) is recovered by
application of the transformation  (\ref{ab}) one obtains the
transition matrix $S$, defined by
\begin{equation}
\varphi _{j}\left( t^{\prime\prime}\right) =\sum\limits^{n_{1}+n{ }
 _{2}}_{j^\prime =1}S_{jj^\prime }\varphi _{j^\prime }\left(
-t^\prime \right)  ,\qquad 1\le j\le n_{1}+n_{2} ,
\end{equation}
in the zero-order approximation, as \begin{mathletters}
\label{Sor}
\begin{eqnarray}
&&S_{jj^\prime }=\sum\limits^{n}_{l=1}X^{*}_{lj}S^{\left( l\right)
 }_{aa}X_{lj^\prime } \nonumber
\\
&&+\sum\limits^{n{ } _{1}}_{l=n+1}X^{*}_{lj}X_{lj}\exp\left( -i
 V^{\left( a\right) }_{ll}\left( t^\prime +t^{\prime\prime}\right)
 \right)  , \label{Soraa}
\\
&&S_{n_1+k,n_1+k^\prime }=\sum\limits^{n}_{l=1}Y^{*}_{lk}S^{\left(
 l\right) }_{bb}Y_{lk^\prime }+\sum\limits^{n{ } _{2}}_{l=n+1}Y^{
*}_{lk}Y_{lk^\prime } \nonumber
\\
&&\times \exp\left( -i V^{\left( b\right) }_{ll}\left( t^\prime
+t^{\prime\prime}\right) -i\beta \left( t^{\prime\prime}{}^{2}
-t^\prime {}^{2}\right) /2\right)  , \label{Sorbb}
\\
&&S_{j,n_1+k^\prime }=\sum\limits^{n}_{l=1}X^{*}_{lj}S^{\left(
 l\right) }_{ab}Y_{lk^\prime } , \label{Sorab}
\\
&&S_{n_1+k,j^\prime }=\sum\limits^{n}_{l=1}Y^{*}_{lk}S^{\left(
 l\right) }_{ba}X_{lj^\prime } , \label{Sorba}
\end{eqnarray}
\end{mathletters} where $1\le j\le n_{1}$, $1\le j^\prime \le n_{1}$,
 $1\le k\le n_{2}$, and $1\le k^\prime \le n_{2}$.
This solution constitutes the quasidegeneracy approximation.

Whenever $S^{\left( l\right) }_{aa}$  or $S^{\left( l\right) }_{bb}$
 are $l$-independent, and $n=n_{1}$  (or $n=n_{2}$),
transitions between states within the corresponding set of $n_{1}$
(or $n_{2}$)
parallel potentials become forbidden due to the unitarity of matrices
$X_{lj}$  and $Y_{jk}$. Such an effect may take place if the
 couplings $g_{l}$  are
close in magnitude or very small. If $n<n_{1}$  (or $n<n_{2}$) such
 transitions
vanish only at low couplings, in which case $|S^{\left( l\right)
 }_{aa}|$ (or $|S^{\left( l\right) }_{bb}|$) are
close to unity.

\section{Applicability criteria} \label{SecCrit}

The quasidegeneracy approximation described in Sec.\
\ref{SecTran} is applicable when the terms neglected in Eqs.\
(\ref{zode}) yield sufficiently small contributions to the transition
amplitudes. First-order perturbation theory estimates these
contributions as
\begin{equation}
\Delta S^{\left( a\right) }_{ll^\prime
 }=\int\limits^{t^{\prime\prime}}_{-t^\prime }a^{\left( 0\right)
*}_{l}\left( t\right) V^{\left( a\right) }_{ll^\prime }a^{\left(
 0\right) }_{l^\prime }\left( t\right) dt , \label{DS}
\end{equation}
and analogous expressions for $\Delta S^{\left( b\right) }_{ll^\prime
 }$, obtained by replacing $a$ with
$b$ everywhere in Eq.\ (\ref{DS}).

An overestimate for these amplitudes can be obtained by
substituting $a^{\left( 0\right) }_{l}\left( t\right) =b^{\left(
 0\right) }_{l}\left( t\right) =1$, resulting in the criteria
\begin{equation}
\left( t^\prime +t^{\prime\prime}\right) \Delta V_{1,2}\ll 1 ,
 \label{smallg}
\end{equation}
where the bandwidths of the potential sets $\Delta V_{1}$  and
 $\Delta V_{2}$  are
defined by Eq.\ (\ref{DV12}). However, in certain situations less
stringent criteria may exist, as can be shown by the use of
approximate expressions for the unperturbed wavefunctions [solutions
of Eqs.\ (\ref{zode})].

Such approximate expressions can be obtained in two limiting
cases. The first one is the asymptotic case, in which the bounds $
-t^\prime $
and $t^{\prime\prime}$ lie far outside the two-state transition
 ranges $g_{l}/\beta $, i. e.,
\begin{equation}
t^\prime +t_{l}\gg g_{l}/\beta  ,\qquad t^{\prime\prime}-t_{l}\gg
 g_{l}/\beta \qquad \left( \text{for all }l\right) .
\end{equation}
In this case, an asymptotic expansion of the confluent
hypergeometric function (see \cite{Abramovitz}) on the  the left-hand
asymptote $t^\prime +t_{l}>-t+t_{l}\gg g_{l}/\beta $ yields
\begin{eqnarray}
a^{\left( 0\right) }_{l}\left( t\right) \approx &&a^{\left( 0\right)
 }_{l}\left( -t^\prime \right) \left( |t|/t^\prime \right) ^{i\lambda
 _l}\exp\left( -i V^{\left( a\right) }_{ll}\left( t^\prime +t\right)
 \right)  , \nonumber
\\
&&{} \label{asn}
\\
b^{\left( 0\right) }_{l}\left( t\right) \approx &&b^{\left( 0\right)
 }_{l}\left( -t^\prime \right) \left( |t|/t^\prime \right) ^{
-i\lambda _l} \nonumber
\\
&&\times \exp\left( -i V^{\left( b\right) }_{ll}\left( t^\prime
+t\right) -i\beta \left( t^{2}-t^\prime {}^{2}\right) /2\right)  ,
 \nonumber
\end{eqnarray}
and on the right-hand asymptote $g_{l}/\beta \ll t-t_{l}
<t^{\prime\prime}-t_{l}$  it yields
\begin{eqnarray}
a^{\left( 0\right) }_{l}\left( t\right) \approx &&a^{\left( 0\right)
 }_{l}\left( t^{\prime\prime}\right) \left( t/t^{\prime\prime}\right)
 ^{i\lambda _l}\exp\left( -i V^{\left( a\right) }_{ll}\left( t
-t^{\prime\prime}\right) \right)  \nonumber
\\
&&{} \label{asp}
\\
b^{\left( 0\right) }_{l}\left( t\right) \approx &&b^{\left( 0\right)
 }_{l}\left( t^{\prime\prime}\right) \left( t/t^{\prime\prime}\right)
 ^{-i\lambda _l} \nonumber
\\
&&\times \exp\left( -i V^{\left( b\right) }_{ll}\left( t
-t^{\prime\prime}\right) -i\beta \left( t^{2}
-t^{\prime\prime}{}^{2}\right) /2\right)  . \nonumber
\end{eqnarray}
Whenever $l>n$, Eqs.\ (\ref{asn}) and (\ref{asp}) become exact [see
Eqs.\ (\ref{abu})]. Hereafter one should set $\lambda _{l}=0$ if $l
>n$.

The first-order corrections to the amplitudes (\ref{DS}) can be
therefore estimated as
\begin{eqnarray}
\Delta S^{\left( a\right) }_{ll^\prime }\approx {V{ } ^{\left(
 a\right) }_{ll^\prime }\over 1+i\lambda _{l^\prime }-i\lambda { }
 _{l}}\Bigl\lbrack &&t^\prime a^{\left( 0\right) *}_{l}\left(
-t^\prime \right) a^{\left( 0\right) }_{l^\prime }\left( -t^\prime
 \right)  \nonumber
\\
&&+t^{\prime\prime}a^{\left( 0\right) *}_{l}\left(
 t^{\prime\prime}\right) a^{\left( 0\right) }_{l^\prime }\left(
 t^{\prime\prime}\right) \Bigr\rbrack  \label{fota} ,
\end{eqnarray}
for the horizontal set, and a similar expression, with $b$
replacing $a$, for the slanted set.

Finally, using Eq.\ (\ref{Vnd}) one can write the applicability
criteria in the form
\begin{equation}
\left( t^\prime +t^{\prime\prime}\right) \Delta V_{1,2}\ll |1
+i\lambda _{l^\prime }-i\lambda _{l}| . \label{bigt}
\end{equation}
Let us consider now the second limiting case, in which both
boundaries  $-t^\prime $ and $t^{\prime\prime}$ lie way inside the
 two-state transition ranges
$g_{l}/\beta $, i. e.,
\begin{equation}
t^\prime +t_{l}\ll g_{l}/\beta  ,\qquad t^{\prime\prime}-t_{l}\ll
 g_{l}/\beta \qquad \left( \text{for all }l\right)  . \label{instr}
\end{equation}
In addition, let $\lambda _{l}\gg 1$, in order to obtain an adiabatic
 evolution.
Within the range defined by Eq.\ (\ref{instr}), the adiabatic energies
are approximately $V^{\left( a\right) }_{ll}\pm g_{l}$, and
\begin{eqnarray}
\left(\begin{array}{c}a^{\left( 0\right) }_{l}\left( t\right)  \\
 b^{\left( 0\right) }_{l}\left( t\right) \end{array}\right) \approx
 &&{a^{\left( 0\right) }_{l}\left( -t^\prime \right) +b^{\left(
 0\right) }_{l}\left( -t^\prime \right) \over 2} \nonumber
\\
&&\times \exp\left( -i\left( V^{\left( a\right) }_{ll}+g_{l}\right)
 \left( t^\prime +t\right) \right)  \nonumber
\\
&&\pm {a^{\left( 0\right) }_{l}\left( -t^\prime \right) -b^{\left(
 0\right) }_{l}\left( -t^\prime \right) \over 2} \nonumber
\\
&&\times \exp\left( -i\left( V^{\left( a\right) }_{ll}-g_{l}\right)
 \left( t^\prime +t\right) \right) \text{  .  }\label{abad}
\end{eqnarray}
Substitution of Eqs.\ (\ref{abad}) in Eq.\ (\ref{DS}), taking
into account Eq.\ (\ref{Vnd}), gives the applicability criteria
\begin{equation}
\left( t^\prime +t^{\prime\prime}\right) \Delta V_{1,2}\ll 1+|g_{l}
-g_{l^\prime }|\left( t^\prime +t^{\prime\prime}\right)  .
 \label{bigg}
\end{equation}
Criteria combining the cases (\ref{smallg}), (\ref{bigt}), and
(\ref{bigg}) can be written with the help of Eq.\ (\ref{lambda}) as
the single expression
\begin{equation}
\left( t^\prime +t^{\prime\prime}\right) \Delta V_{1,2}\ll 1+|g_{l}
-g_{l^\prime }|\min\left( t^\prime +t^{\prime\prime},\left( g_{l}
+g_{l^\prime }\right) /\beta \right)  . \label{AC}
\end{equation}
These criteria allow for an interpretation that stems from the
viewpoint of the uncertainty principle. Equation  (\ref{AC}) means
that the potentials become indistinguishable within a limited time
interval. The second term in the right-hand side of Eq.\ (\ref{AC})
describes a broadening of the allowed uncertainty as the coupling
increases.

\section{Results and discussion} \label{SecDisc}

In the limiting case of a linear grid defined on the infinite
time interval $-\infty <t<\infty $, some transitions become forbidden
 (see
\cite{BE93}). An example of such transitions is shown in Fig.\
\ref{FigPot}, in which two time-independent potentials are shown
crossed by three parallel time-slanted potentials. The forbidden
transitions, such as $2\rightarrow 1$, $3\rightarrow 4$,
 $3\rightarrow 5$, and $4\rightarrow 5$, are called
counterintuitive, since in order to treat them as a sequence of
independent two-state crossings, one has to assume a motion backwards
in time.

Counterintuitive transitions can nonetheless occur, as has been
proven in numerical calculations involving crossings of nonlinear
potentials \cite{NWJ97}, and in uses of the quasidegeneracy
approximation for truncated and piecewise linear problems, involving a
set of horizontal potentials, crossed by one slanted potential
\cite{YBJB99}. Such transitions are present in truncated linear grids
as well, since the transformations (\ref{ab}) connect the initial and
final states to all the decoupled channels.

Hereafter we shall demonstrate the application of the
quasidegeneracy approximation to a particular example. Consider the
model of a linear grid  with $n_{1}=n_{2}=2$, $V_{1}=V_{3}=-\Delta V
/2$, and $V_{2}=V_{4}=\Delta V/2$
(recalling that $V_{3}$  and $V_{4}$  are the time-independent parts
 of the
slanted potentials). Let the coupling matrix have one of the two
special forms, either
\begin{equation}
g_{jk}=g_{0}\left(\begin{array}{cc}1/1.2&1 \\
 1&1.2\exp\left( i m\pi /4\right) \end{array}\right)  , \label{g2}
\end{equation}
with integer values of $m$, or the equal-coupling form, with
\begin{equation}
g_{11}=g_{12}=g_{21}=g_{22}=g_{0} . \label{geq}
\end{equation}
All the following calculations are performed for the slope $\beta =1$.
The results can be readily expanded to other $\beta $ values by the
substitutions $g/\sqrt{\beta }\rightarrow g$, $\Delta V/\sqrt{\beta
 }\rightarrow \Delta V$, and $t\sqrt{\beta }\rightarrow t$.

\begin{figure}
\psfig{clip=,figure=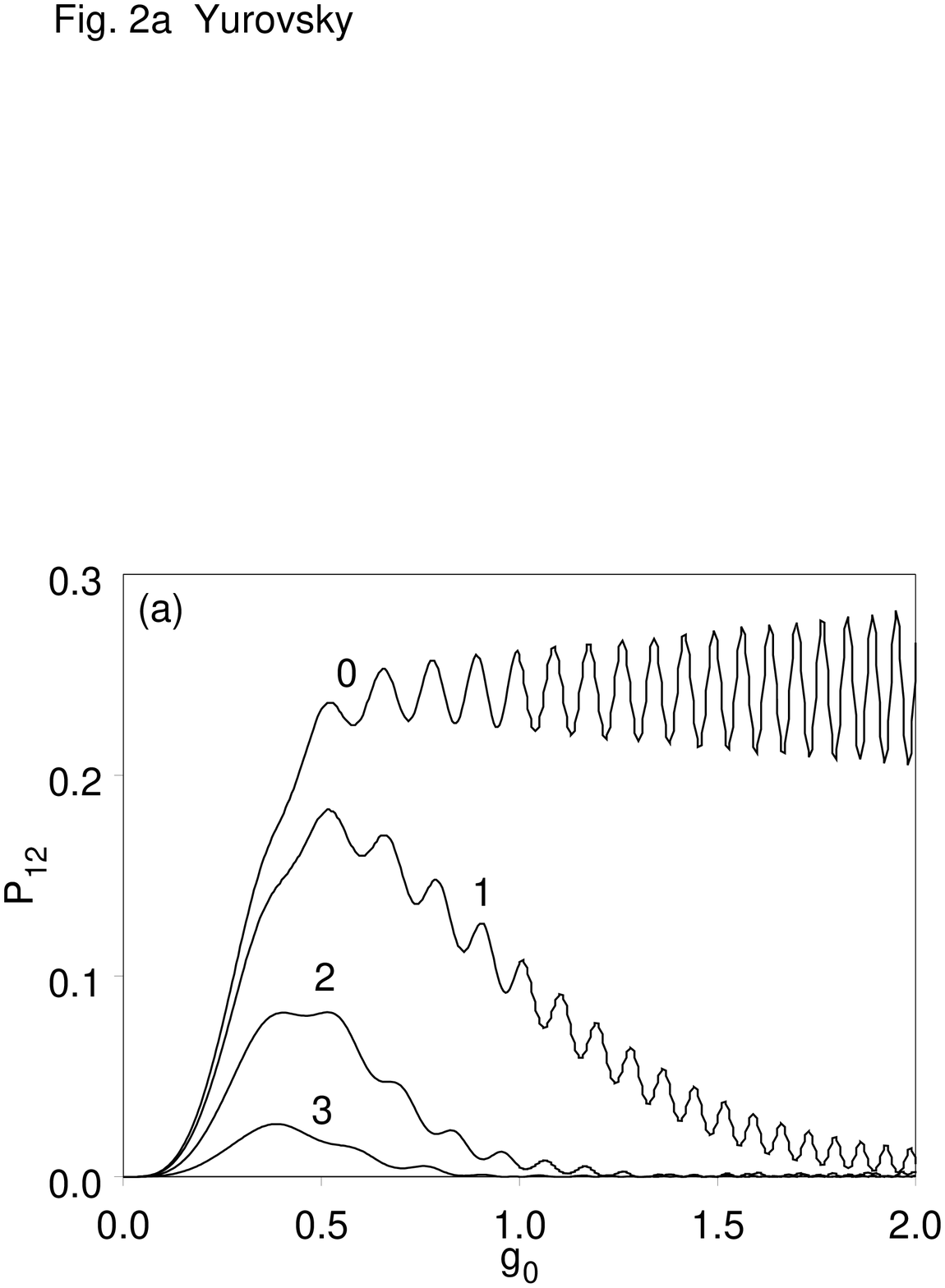,width=3.375in}
\psfig{clip=,figure=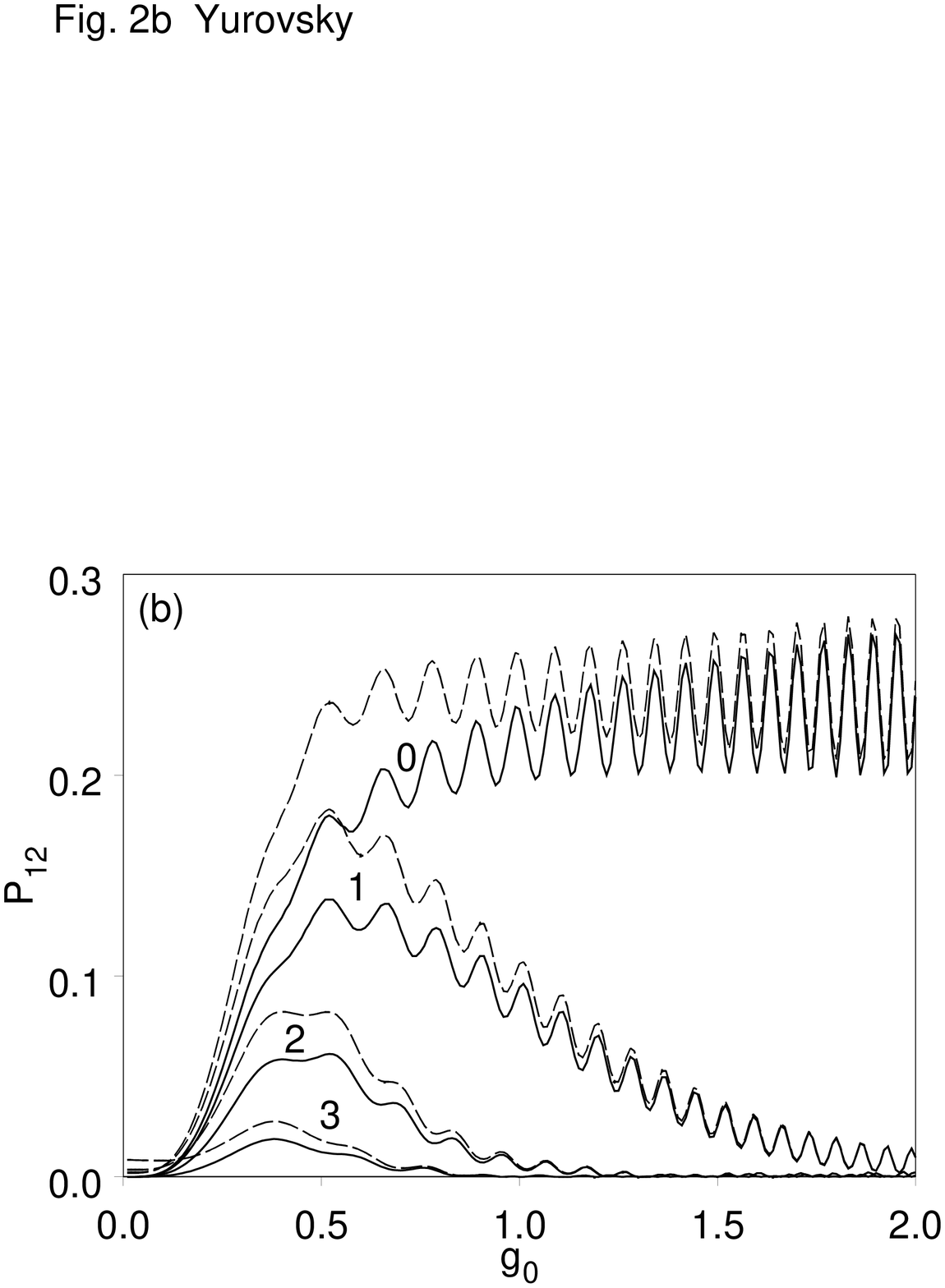,width=3.375in}

\caption{Counterintuitive transition probabilities vs. the
coupling strength $g_{0}$ [see Eq.\ (\protect\ref{g2})] for a truncated
linear grid with the bounds $t^\prime =t^{\prime\prime}=100$ (on a
 scale in which the
potential slopes $\beta =1$) and the potential gaps (a) $\Delta V=0$,
 and (b)
$\Delta V=2.5\times 10^{-3}$. The numbers denote the values of the
 phase parameter $m$ in
Eq.\ (\protect\ref{g2}). The results of numerical integration of the
coupled equations (\protect\ref{phieq}) are presented by solid lines.
The dashed-line plots in part (b) are calculated with the
quasidegeneracy approximation using Eqs.\ (\protect\ref{Sor}).}
\label{FigCoup}

\end{figure}

Figure \ref{FigCoup} presents the dependence of counterintuitive
transition probabilities on the coupling strength $g_{0}$  for two
 cases:
an exactly degenerate one ($\Delta V=0$), and a one in which $\Delta
 V\left( t^\prime +t^{\prime\prime}\right) =0.5$, on
the verge of the validity criteria (\ref{AC}). At low values of
 $g_{0}$
the amplitudes $S^{\left( l\right) }_{aa}$  and $S^{\left( l\right)
 }_{bb}$  in the decoupled representation Eq.\
(\ref{Sl}) are close to unity and practically independent of $l$, and
therefore all transitions (including counterintuitive ones) within
each of the two sets of parallel potentials in the original
representation have small probabilities [see discussion following Eqs\
(\ref{Sor})]. In cases in which the singular values are quite similar,
the probabilities of such transitions become small at high coupling
strengths, since $S^{\left( l\right) }_{aa}$  and $S^{\left( l\right)
 }_{bb}$  are small for all $l$ (see, for
example, the plots for $m=3$ in Fig.\ \ref{FigCoup}, where
 $g_{1}=1.73g_{0}$
and $g_{2}=1.07g_{0}$).

However, if the singular values of the coupling matrix are
significantly different, the counterintuitive transitions remain
significant over a wide range of coupling strengths as some of  the
amplitudes $S^{\left( l\right) }_{aa}$  (or $S^{\left( l\right)
 }_{bb}$) are large, and some are small (see the
plots for $m=1$ in Fig.\ \ref{FigCoup}, where $g_{1}=2g_{0}$  and
 $g_{2}=0.38g_{0}$). In
the case of a separable matrix (see the plot for $m=0$ in Fig.\
\ref{FigCoup}, where $g_{1}=2.03g_{0}$  and $g_{2}=0$), such
 transitions persist
even in the limit of high coupling strength. It is worth noting that
even a change of the phase of one element of the coupling matrix
transforms a separable matrix to a non-separable one, and therefore
changes the behavior of the transition probability at high values of
the coupling strengths.

\begin{figure}
\psfig{clip=,figure=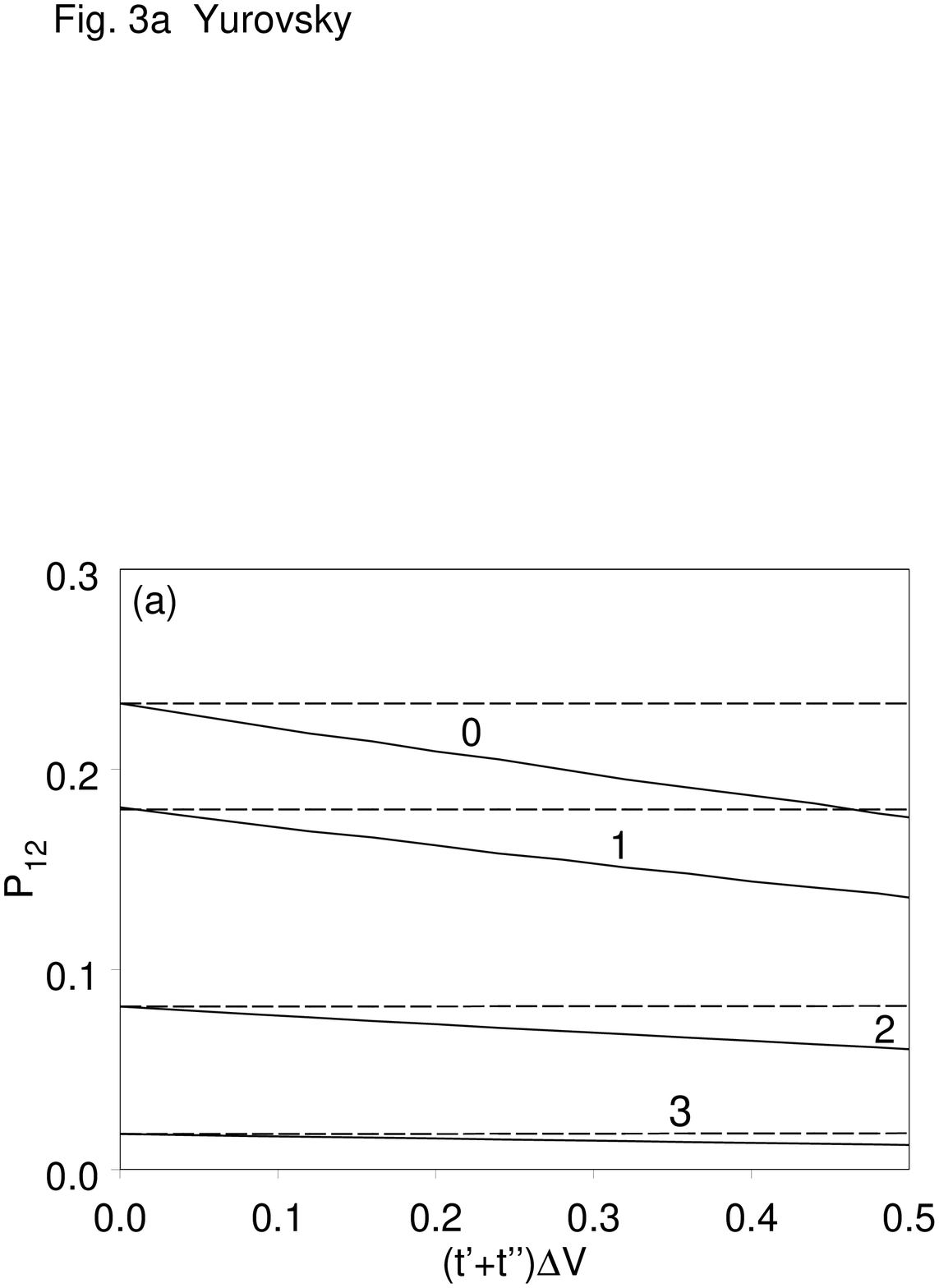,width=3.375in}
\psfig{clip=,figure=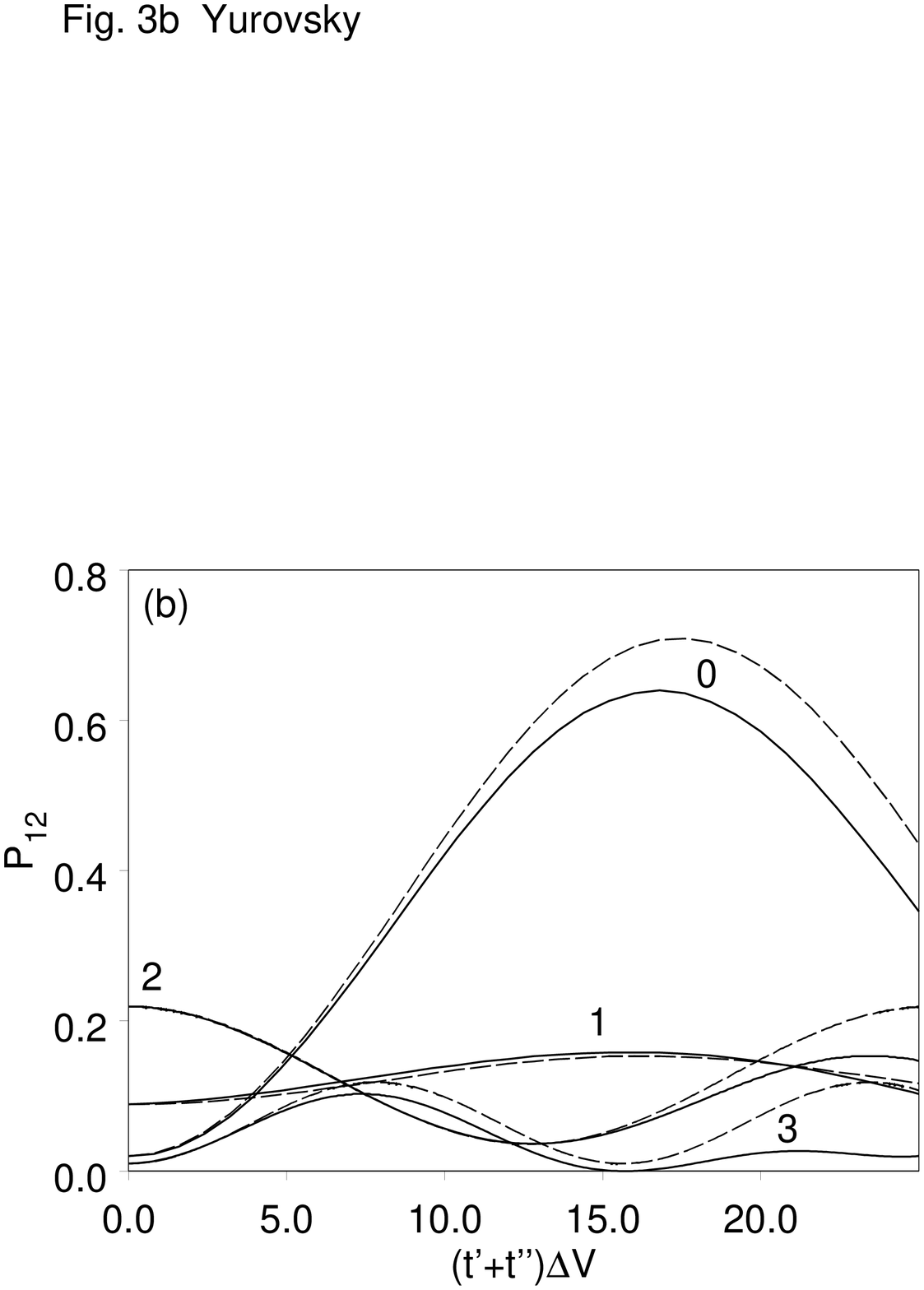,width=3.375in}

\caption{Counterintuitive transition probabilities vs. the
potential gap $\Delta V$, calculated for (a) $t^\prime
 =t^{\prime\prime}=100$, $g_{0}=0.5$,  or (b)
$t^\prime =t^{\prime\prime}=20, g_{0}=5$. Other notations as in Fig.\
 \protect\ref{FigCoup}.}
\label{FigGap}

\end{figure}

Counterintuitive transitions persist at finite values of the
potential gap $\Delta V$ as well (see Figs.\ \ref{FigCoup}b and
 \ref{FigGap}).
As one can see, the higher is the coupling strength, the better are
the results of the quasidegeneracy approximation [in agreement with
the criteria Eq.\ (\ref{AC})]. At low coupling strengths  the
predictions of the quasidegeneracy approximation are correct as long
as $\left( t^\prime +t^{\prime\prime}\right) \Delta V\le 0.2$ (see
 \ref{FigGap}a), while at high coupling strengths
they are correct as long as $\left( t^\prime +t^{\prime\prime}\right)
 \Delta V\le 0.2\lambda $ (see \ref{FigGap}b).

\begin{figure}
\psfig{clip=,figure=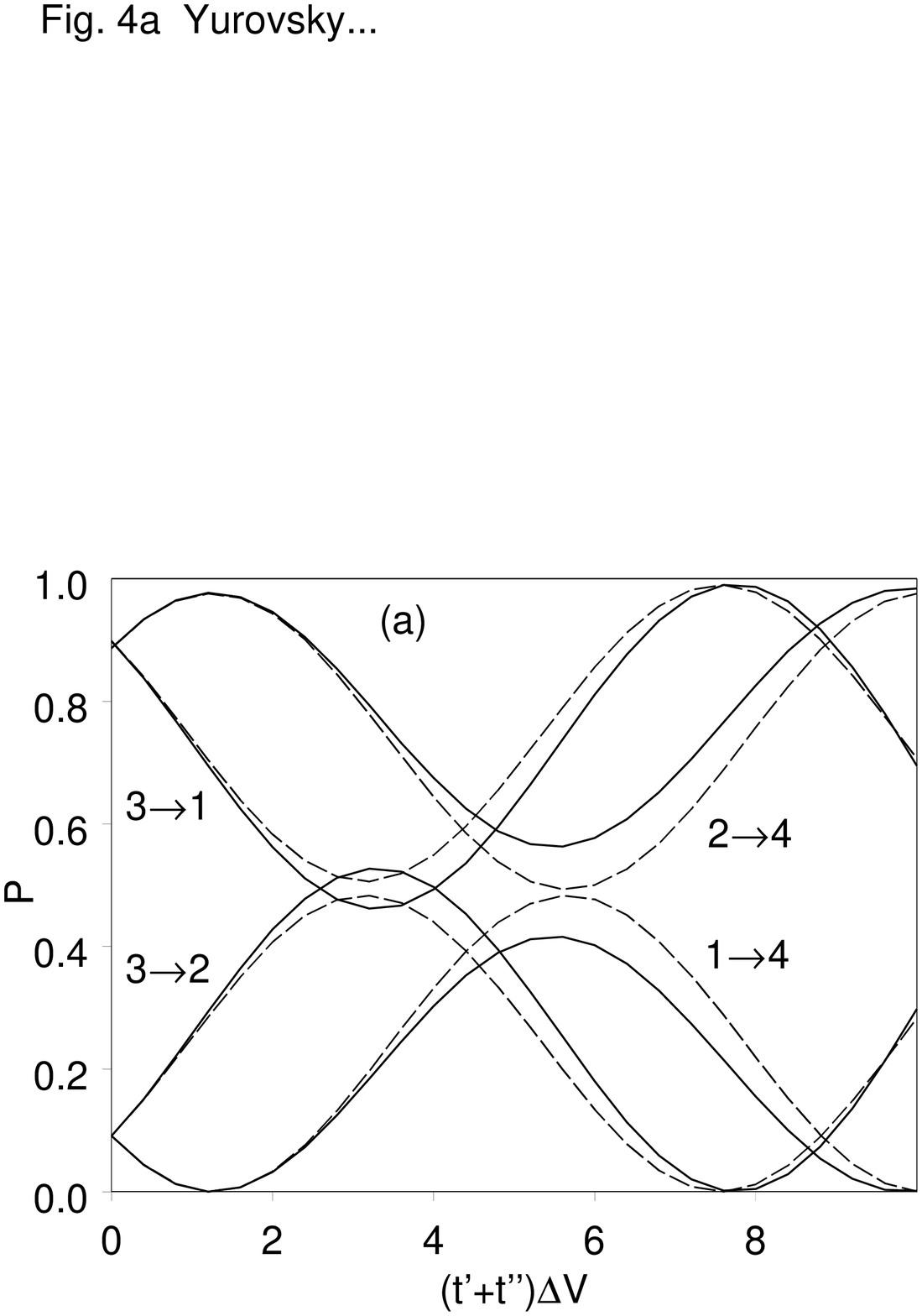,width=3.375in}
\psfig{clip=,figure=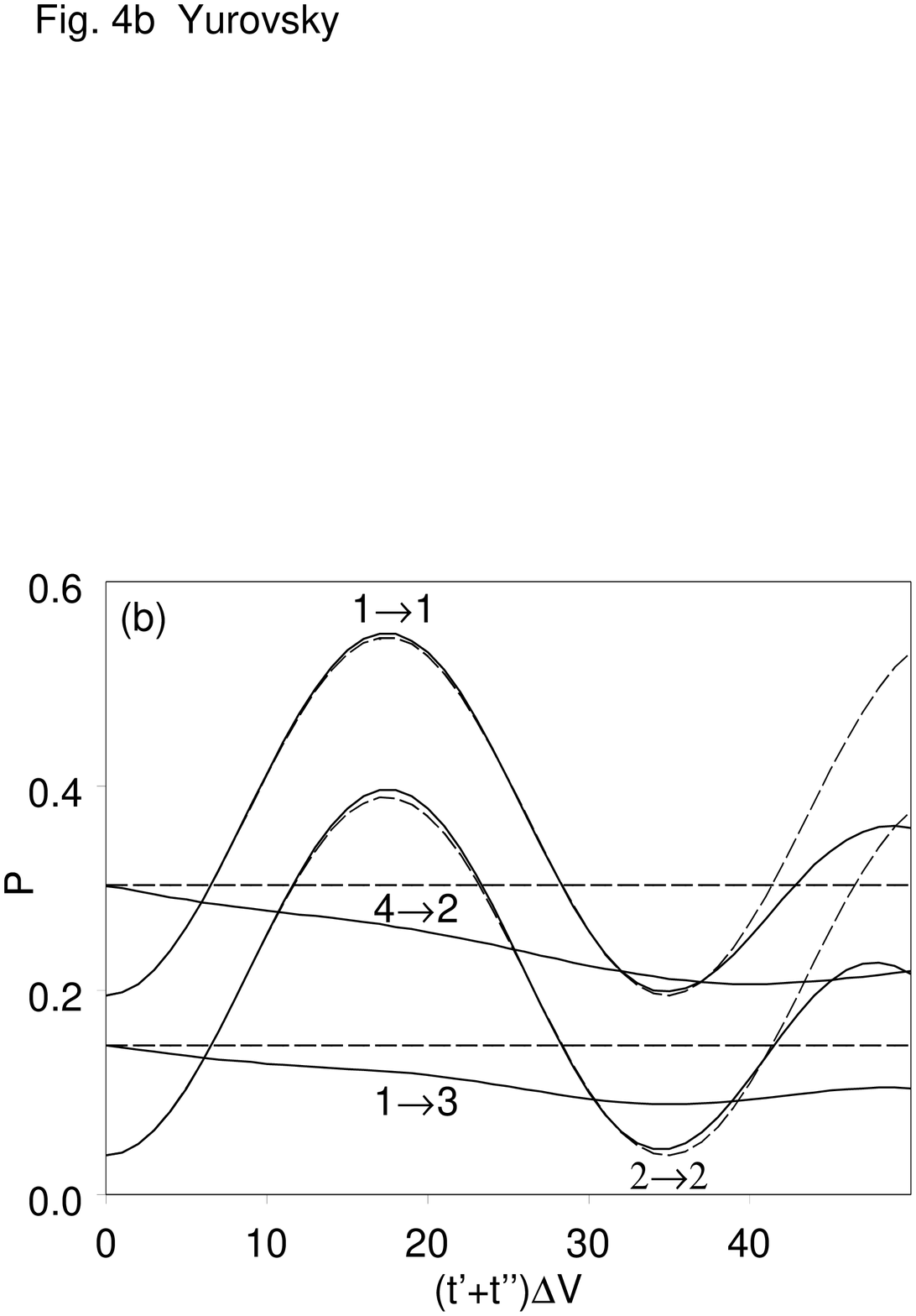,width=3.375in}
\psfig{clip=,figure=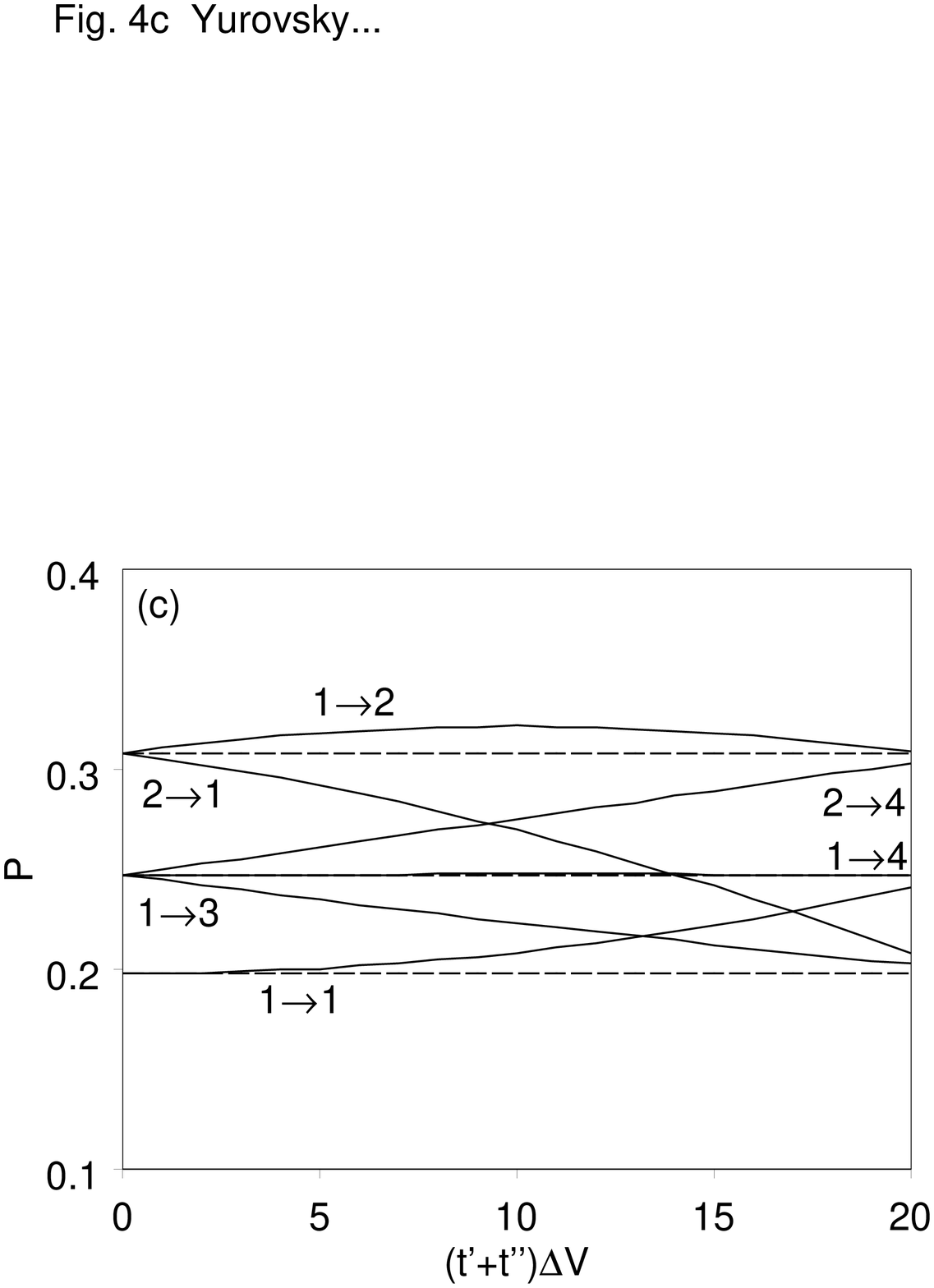,width=3.375in}

\caption{Probabilities of specified state-to-state transitions vs. the
potential gap $\Delta V$ for a truncated linear grid with $t^\prime
 =t^{\prime\prime}=50$ and $g_{0}=5$. Parts
(a) and (b) correspond to the coupling matrix (\protect\ref{g2}) with
 $m=4$
and $m=0$, respectively, while part (c) corresponds to the case of
 equal
couplings (see Eq.\ (\protect\ref{geq}). Other notations as in Fig.\
\protect\ref{FigCoup}.} \label{FigOsc}

\end{figure}

Probabilities of counterintuitive transitions (see Fig.\
\ref{FigGap}b) and other transitions (see Fig.\ \ref{FigOsc})
demonstrate an oscillating pattern in their dependence on the
potential gap. The nature of these oscillations is different from the
well-known St\"uckelberg oscillations (see Ref.\ \cite{ChildSM}),
 which
may be present only in transitions including two or more interfering
``intuitive'' paths. (Such paths exist in the transitions from 1 or 4
to 2 or 3 in the case presented in Figs.\ \ref{FigGap} and
\ref{FigOsc}.) The period of the St\"uckelberg oscillations is
 $\Delta V/\beta $;
i.e., it is dependent on the potential gap $\Delta V$ but independent
 of the
time interval $t^\prime +t^{\prime\prime}$. These properties, as well
 as the magnitude of the
St\"uckelberg oscillation period, are not in agreement with the
 behavior
of the oscillations presented in Figs.\ \ref{FigGap}b and
\ref{FigOsc}.

The quasidegeneracy approximation relates the oscillations
reported here to the interference of the terms in Eqs.\ (\ref{Sor}),
corresponding to different decoupled channels. The dependence on
 $\Delta V$ is
due to exponents in Eqs.\ (\ref{fundsol}) and in the second sum of
each of the two equations (\ref{Soraa}) and (\ref{Sorbb}). The
oscillation period in $\Delta V$ is $2\pi \rho /\left( t^\prime
+t^{\prime\prime}\right) $, where $\rho =\Delta V/\left( V^{\left(
 a\right) }_{22}-V^{\left( a\right) }_{11}\right)
=\Delta V/\left( V^{\left( b\right) }_{22}-V^{\left( b\right)
 }_{11}\right) $ is the ratio of the potential gaps in the original
 and
decoupled representations. For the coupling matrix (\ref{g2}) we have
$\rho =5.6$, 5.3, 4.0, and 2.5 for $m=0,1,2$, and 3, respectively,
 which
explains the variation of the oscillation period with $m$ in Fig.\
\ref{FigGap}. It is worth noting that in Fig.\ \ref{FigOsc}b  these
oscillations are absent just for the transitions for which one would
expect St\"uckelberg oscillations ($4\rightarrow 2$ and $1\rightarrow
 3$). The reason for not
seeing St\"uckelberg oscillations in our figures is simple. The scale
 of
the plot is too small to show even a single St\"uckelberg  period. In
the case of equal coupling [see Eq.\ (\ref{geq})] $V^{\left( a\right)
 }_{ll}=V^{\left( b\right) }_{ll}=0$ for all
$l$. This property results in the absence of oscillations in Fig.\
\ref{FigOsc}c.

There is still another kind of oscillations possible. It has been
demonstrated in the case of a truncated two-state linear curve
crossing, in which oscillations may show up as a function of each of
the two truncation times  (see \cite{VG96}). In principle, such
oscillations should also appear in our model in the $\Delta V$
 dependence,
since the crossing points move as the potential gap is varied.
However, the period of these oscillations, too, is too large to show
up in Figs.\ \ref{FigGap} and \ref{FigOsc}.

In the limit of high coupling strengths or slow potential
variation ($\lambda \gg 1$) the semiclassical approach of independent
 crossings
predicts non-vanishing transitions only for one final state per a
given initial state. If $n_{1}=n_{2}$  all the non-vanishing
 transitions lead
from one set of the parallel potentials to another set, leading to a
complete population transfer between the sets. This property was used
in a recent proposal of single-electronics devices, based on
transitions between quantum dots (Refs.\ \cite{U97,U98}). In contrast,
the quasidegeneracy approximation predicts more non-vanishing
transitions (see Figs.\ \ref{FigGap} and \ref{FigOsc}). In the case of
$n<n_{1}=n_{2}$,  a finite probability may remain for transitions
 within the
same set of the parallel potentials (see Figs.\ \ref{FigGap},
\ref{FigOsc}b, and \ref{FigOsc}c), leading to an incomplete population
transfer. This effect is similar to the effect of incomplete optical
shielding in ultracold atom collisions \cite{NWJ97,YB97}. The effect
of incomplete population transfer may interfere with the operation of
the single-electronics devices mentioned above.

\section{Conclusions}

Equations (\ref{Sor}) describe the transition amplitudes in a
truncated linear potential grid, whenever the applicability criteria
(\ref{AC}) are observed. The results can be applied also to a more
general case, in which the grid may be broken into well-separated
groups of quasidegenerate crossings. In this case the transition
amplitudes can be represented as products of the transition amplitudes
given by Eqs.\ (\ref{Sor}) for the quasidegenerate groups. Thus the
approximation can be used in a wide variety of physical problems in
which a multistate curve crossing occurs.

\newpage
\newpage
\newpage
\end{document}